\documentclass[aps, prd, twocolumn, lengthcheck, superscriptaddress, showpacs, letterpaper, nofootinbib]{revtex4-1}

\usepackage{amsmath,amssymb}
\usepackage{graphicx,bm}
\usepackage{slashed}
\usepackage{dcolumn}
\usepackage{epstopdf}
\usepackage{ulem} 
\usepackage[usenames]{color}
\usepackage{mathrsfs}

\allowdisplaybreaks

\def\la{\langle}\def\ra{\rangle}
\def\be{\begin{eqnarray}}\def\ee{\end{eqnarray}}
\def\lsim{\mathrel{\rlap{\lower3pt\hbox{\hskip1pt$\sim$}}
     \raise1pt\hbox{$<$}}} 
\def\gsim{\mathrel{\rlap{\lower3pt\hbox{\hskip1pt$\sim$}}
     \raise1pt\hbox{$>$}}} 
\def\le{ \begin{array}{ll}}\def\re{\end{array}}

\def\lear{ \left( \begin{array}{cc}}\def\rear{\end{array} \right)}

\def\le{ \left( \begin{array}{cc}}\def\re{\end{array} \right)}

\def\bi{\bibitem}

\def\eft-hls{{\it EFT}$_{\rm bsHLS}$}
\def\skyrmion-hls{{\it Skyrmion}$_{\rm sHLS}$}

\def\Lamb-chiral{\Lambda_{\rm chiral}}
\def\Lamb-bshls{\Lambda-{\rm bshls}}

\def\Esym{E_{\rm sym}}
\def\del{\partial}

\begin{document}

\title{Cusp in the Symmetry Energy,  
Speed of Sound in Neutron Stars \\ and Emergent Pseudo-Conformal Symmetry }

\author{Hyun Kyu Lee}
\email{hyunkyu@hanyang.ac.kr}
\affiliation{Department of Physics, Hanyang University, Seoul 133-791, Korea}

\author{Yong-Liang Ma}
\email{ylma@ucas.ac.cn}
\affiliation{School of Fundamental Physics and Mathematical Sciences,
Hangzhou Institute for Advanced Study, UCAS, Hangzhou, 310024, China}
\affiliation{International Center for Theoretical Physics Asia-Pacific, Beijing/Hanzhou, China}

\author{Won-Gi Paeng}
\email{wgpaeng@clunix.com}
\affiliation{A I Lab., Clunix, Seoul 07209, Korea}


\author{Mannque Rho}
\email{mannque.rho@ipht.fr}
\affiliation{Universit\'e Paris-Saclay, Institut de Physique Th\'eorique, CNRS, CEA,  91191, Gif-sur-Yvette, France }

\date{\today}
\begin{abstract}
We review how the ``cusp" predicted in the nuclear symmetry energy generated by a topology change at density $n_{1/2}\gsim 2 n_0$ can have a surprising consequence, so far unrecognized in nuclear physics and astrophysics communities,  on the structure of dense compact-star matter. The topology change, when translated into nuclear EFT with ``effective"  QCD degrees of freedom in terms of hidden local and scale symmetries duly taken into account,  predicts an EoS that is soft below and stiff above  $n\gsim n_{1/2}$, involving no low-order phase transitions, and yields the macrophysical properties of neutron stars consistent -- so far with no tension --  with the astrophysical observations, including  the maximum mass $ 2.0\lsim M/ M_\odot\lsim 2.2$ as well as the GW data.  Furthermore it describes the interior core of the massive stars populated by baryon-charge-fractionalized quasi-fermions that are neither baryonic nor quarkonic.  It is argued that the cusp ``buried" in the symmetry energy resulting from strong correlations with hidden heavy degrees of freedom leads, at $n\gsim n_{1/2}$, to what we dubbed ``pseudo-conformal" sound speed,  $v^2_{pcs}/c^2\approx 1/3$,  precociously converged from below at $n_{1/2}$. It is not strictly conformal since  the trace of energy-momentum tensor is not zero even in the chiral limit. This observation with the topology change identified with the putative hadron-quark continuity, taking place at at density $\gsim 2 n_0$, implies that the quantities accurately measured at $\sim n_0$ cannot give a stringent constraint for what takes place at the core density of compact stars $\sim (3-7) n_0$. This is because the change of degrees of freedom in effective field theory is involved. We discuss the implication of this on the recent PREX-II ``dilemma" in the measured skin thickness of $^{208}$Pb.
\end{abstract}


\maketitle


\section{introduction}
%
In accessing  dense neutron-star matter in terms of a topology change for the putative hadron-quark continuity, it was discovered  in 2011~\cite{LPR-cusp} that a cusp is present in the nuclear symmetry energy $\Esym$ at a density $\sim (2-3)$ times the normal nuclear matter density $n=n_0\simeq 0.16$ fm$^{-3}$. This cusp structure has been found to play the  most  important role in the approach to the EoS of dense compact-star matter developed entirely independently of other on-going approaches in nuclear astrophysics.
Formulated with the minimum number of degrees of freedom available it has the power to go beyond the standard chiral effective field theory (S$\chi$EFT), currently heralded as  as a possible ``first-principles approach" to nuclear theory at low energy and density, and gives extremely simple predictions that have the merit to be unambiguously confronted by experiments in the density regime inaccessible by S$\chi$EFT. It has thus far accounted with no unsurmountable tension for all macro-physical observables available in both terrestrial and astrophysical laboratories. See for the current status, e.g., \cite{MR-review}.

 In this paper, we show that this cusp structure zeroes in on the recent issue raised by the  PREX-II measurement of the neutron skin thickness of $^{208}$Pb~\cite{PREX-II} and the impact on the equation of state (EoS) of massive compact stars. An analysis using the new $R_{\rm skin}^{208}$ and certain correlations with  the symmetry energy $J$ and its slope $L$ (to be defined) at $n=n_0$  led to  the 1 $\sigma$ intervals~\cite{reedetal}
\be
J=(38.1\pm 4.7)\ {\rm MeV}, \ L=(106\pm 37)\ {\rm MeV}.\label{nonnormal}
\ee
These values seemingly overshoot  the currently ``accepted" values~\cite{acceptedvalues} \footnote{We will elaborate on these ``accepted" values below},
\be
J=(31.7\pm 1.1)\ {\rm MeV}, \ L=(59.8\pm 4.1)\ {\rm MeV}.\label{accepted}
\ee
This result is taken to imply by some nuclear physicists that the EoS could  be a lot stiffer at normal nuclear matter density -- hence at higher densities -- than what  has been considered up to date. A similar observation termed as a ``dilemma" is arrived at by Piekarewicz  from the electric dipole polarizability of neutron-rich nuclei~\cite{dilemma}. Naively extrapolated to the massive compact-star density,  the $R_{\rm skin}^{208}$ data could rule out most of, or at least put in serious tension, the EoS' currently available in the literature for compact-star physics.  

The stiff EoS implied by the dilemma turns out, as we will discuss later, to have a dramatic effect on the properties of massive stars such as the composition of the star core and sound speed. What we will show is that the cusp structure discovered in \cite{LPR-cusp} gives rise to a drastically different picture. This would revamp the popularly accepted notion in certain nuclear astrophysics circles that the EoS determined accurately at low density, say, at $\sim n_0$, should make an ``indispensable" constraint to the  EoS at higher densities. Put differently,  what we shall refer to as nuclear-astrophysics`` lore" (nLORE for short) states that what happens in the core of compact stars  be constrained by what happens in nuclear matter. This of course must be true in a uniquely given theory, namely, QCD or an effective field theory (EFT) with UV completion. However at present QCD can directly access neither nuclear matter nor compact-star matter and what's available is effective field theory (EFT) in the sense defined by Weinberg's Folk Theorem. In EFT, this nLORE cannot be valid if there are phase changes or crossovers at different scales in density in the present case. In fact, we will argue the presence of the cusp in our approach debunks the nLORE on constraints on EoS. What turns out to importantly figure in our argument is the existence of that cusp  at a density $\gsim  (2-3)n_0$ in the symmetry energy induced by a (robust)  topology change in dense matter that effectively encodes the putative hadron-quark continuity expected in QCD.  It aptly reconciles a soft EoS at $n \lsim n_{1/2}$ to a hard EoS at $n > n_{1/2}$, accounting notably, among others,  for massive $\gsim 2 M_\odot$ compact-stars and other macroscopic star properties including the recent gravity wave data.  
\section{The cusp in $\mathbf{E_{\rm sym}}$}
The quantity that plays the most important role in the EoS for compact-star matter~\cite{Steiner:2004fi} is the symmetry energy $\Esym$ in the energy per nucleon given by
\be
E(n, \alpha) & = & E(n, \alpha = 0) + E_{\rm sym}(n)\alpha^2 + O(\alpha^4) + \cdots ,
\label{eq.Esym}
\ee
where $\alpha=(N-P)/(N+P)$ is the neutron-proton asymmetry with $P$ ($N$) standing for  the number of protons (neutrons) in $A=N+P$ nucleon system. The $J$ and $L$ concerned are
\be
J=\Esym (n_0), \ L=3 \frac{\del \Esym(n)}{\del n}|_{n=n_0}.
\ee
The issue associated with the Pb skin thickness puzzle involves this symmetry energy on which our discussion will be focused. In standard nuclear physics approaches (SNPA)  anchored on effective density functionals  such as the Skyrme potential, relativistic mean field (RMF) and varieties thereof  as well as S$\chi$PT up to a manageable chiral order, typically N$^3$LO, equipped with a certain number of parameters fit to available empirical data, the $E(n,\alpha)$ can be more or less reliably determined in the vicinity of the nuclear matter equilibrium density $n_0$. It has also been extended, with albeit significant uncertainty, up to slightly above $n_0$ from heavy-ion collision experiments.  Thus one can say that the nuclear symmetry energy $\Esym$  is fairly well determined up to  $n_0$ in SNPAs. It should, however, be stressed that its slope in density, namely, $L$  and higher derivatives remain uncertain, say in S$\chi$PT, unless  chiral-order terms up to N$^m$LO for $m\gsim 4$ are fully included.  This  is closely tied to the fact that the chiral power expansion  (say, in S$\chi$EFT) is bound to break down for $k_F^\kappa$ for $\kappa \gsim 5$ ((e.g., \cite{holt-rho-weise}) as the hadron-quark crossover density is approached with change of degrees of freedom. So the problem is how $\Esym$ and its derivatives behave beyond the equilibrium density $n_0$. This is where heavy degrees of freedom (HDsF) can enter.
\subsection{Cusp in Skyrmion Crystal}
We address this problem exploiting a topological structure of dense baryonic matter. This is because in the large $N_c$ limit in QCD, the only known non-perturbative tool available in strong interaction physics applicable to baryonic matter at large density -- in the absence of lattice QCD -- is  putting skyrmions (or instantons in holographic QCD) on crystal lattice~\cite{crystal}. Application of the crystal skyrmion lattice method to nuclear matter and dense matter has been around for some time (see for an early review \cite{PV-lattice}) but only recently is the power of the skyrmion approach beginning to be recognized in nuclear physics,  contrary to condensed matter as well as string theory where the skyrmion structure in various spatial dimensions has been having remarkable impacts~\cite{multifacet}.  This is because of the extreme mathematical subtlety involved in the skyrmion physics. Furthermore, the condition for the validity of lattice skyrmions in particular, i.e. large $N_c$ and large density,  is not met at the density where there is a wealth of experimental data, namely finite nuclei. However the cusp structure in question that takes place at relatively high density -- relative to normal nuclear matter -- seems to meet the two conditions as indicated by the quasi-scale invariance seen in the crystal simulation in the half-skyrmion phase~\cite{PKLMR}.  

To illustrate the basic idea, we first take  the Skyrme model~\cite{Skyrme} stabilized by the (Skyrme) quartic term for skyrmions supplemented by a scalar dilaton as first shown in \cite{LPR-cusp}. What is crucial is that the Skyrme model encodes the necessary topological structure. But by itself, it misses certain nontrivial crucial dynamical characteristics encoded in QCD. We will  implement these missing ingredients with hidden local symmetry (HLS for short) supplemented with hidden scale symmetry (HSS)  and incorporate them for quantitative discussions in the generalized chiral effective field theory (EFT) approach that is dubbed $Gn$EFT~\cite{MR-review}.  We will present the argument that the HLS and HSS (combined, referred to as sHLS),  the degrees of freedom associated with which are identified with the HDsF involved at high density,  are ``dual" to QCD (gluons and quarks)\footnote{This notion of hadron-quark duality will be specified below.} in the density regime relevant to compact stars. It will be argued that the density involved  is located far below asymptotic density at which hardon-quark continuity presumably does break down (to be specified below).

Following  \cite{LPR-cusp}, we calculate the symmetry energy by quantizing  the crystal as a whole object through a collective rotation in iso-space with the rotation angle $C(t)$ acting on the relevant chiral fields $U=\xi^2$ (in unitary gauge) as
\begin{eqnarray}
\xi_c(\mathbf{x}) & \to & \xi(\mathbf{x}, x) = C(t) \xi_c(\mathbf{x}) C^\dagger(t),
\nonumber \\
V_{\mu,c}(\mathbf{x}) & \to & V_{\mu}(\mathbf{x},t) =  C(t) V_{\mu,c}(\mathbf{x}) C^\dagger(t) ,
\label{eq:rotation}
\end{eqnarray}
where the subindex ``$c$" means the static configuration with the lowest energy for a given crystal size $L$ and $C(t)$ is a time-dependent unitary $SU(2)$ matrix in isospace. We  define the angular velocity through ${\bm{\Omega}}$
 \begin{eqnarray}
\frac{i}{2}\bm{\tau}\cdot\bm{\Omega} & = & C^\dagger(t)\partial_0 C(t) .
\label{eq:defvelocity}
\end{eqnarray}
The energy of the $A$-nucleon system can be written as
\begin{eqnarray}
M_{\rm tot} & = & M_{\rm static} + \frac{1}{2}\lambda_{I}^{\rm tot} \bm{\Omega}^2.
\end{eqnarray}
By regarding the angular momentum in isospace, $\mathbf{J} = \delta M_{\rm tot}/\delta \bm{\Omega}$, as the isospin operator, one can write the total energy of the system  as
\begin{eqnarray}
M_{\rm tot} & = & A M_{\rm sol} + \frac{1}{2A \lambda_{I}}I^{\rm tot}(I^{\rm tot} + 1),
\end{eqnarray}
where $M_{\rm sol}$, $\lambda_{I}$ and $I^{\rm tot}$ are, respectively, the mass and moment of inertia of the single skyrmion in the system, and the total isospin of the $A$-nucleon.
Given that the $A$-nucleon system is taken a nearly pure neutron system,  $I^{\rm tot} \leq A/2$, to the leading order of $A$ for $A \to \infty$, the energy per baryon takes the form\footnote{There is in principle the Casimir energy of $O(N_c^0)$ but it does not enter in the symmetry energy.}
\begin{eqnarray}
E & = & M_{\rm sol} + \frac{1}{8 \lambda_{I}}\alpha^2.
\end{eqnarray}
Thus the symmetry energy is
\begin{eqnarray}
E_{\rm sym} & = & \frac{1}{8 \lambda_{I}} +O(1/N_c^2).
\label{eq:LEsymLambda}
\end{eqnarray}
The moment of inertia $\lambda_I\sim O(N_c)$   can be computed in the leading $N_c$ order as the integral over the single cell and takes the form \begin{eqnarray}
\lambda_{I} & = & \frac{f_\pi^2}{6}\left\langle \left(4 - 2 \phi_0^2\right)\right\rangle + \delta\lambda_{I} +\cdots,
\label{lambdaI}
\end{eqnarray}
where the first term comes from the quadratic current algebra term and the second stands for the contribution from the Skyrme quartic term which consists of four terms involving $\phi_0$ and space derivatives of the chiral field $\xi$. Here $\phi_0$, proportional to the quark condensate $\la\bar{q}q\ra$,  plays a crucial role in the whole development in \cite{MR-review}. 

\begin{figure*}[htb]
\centering
\includegraphics[width=10cm]{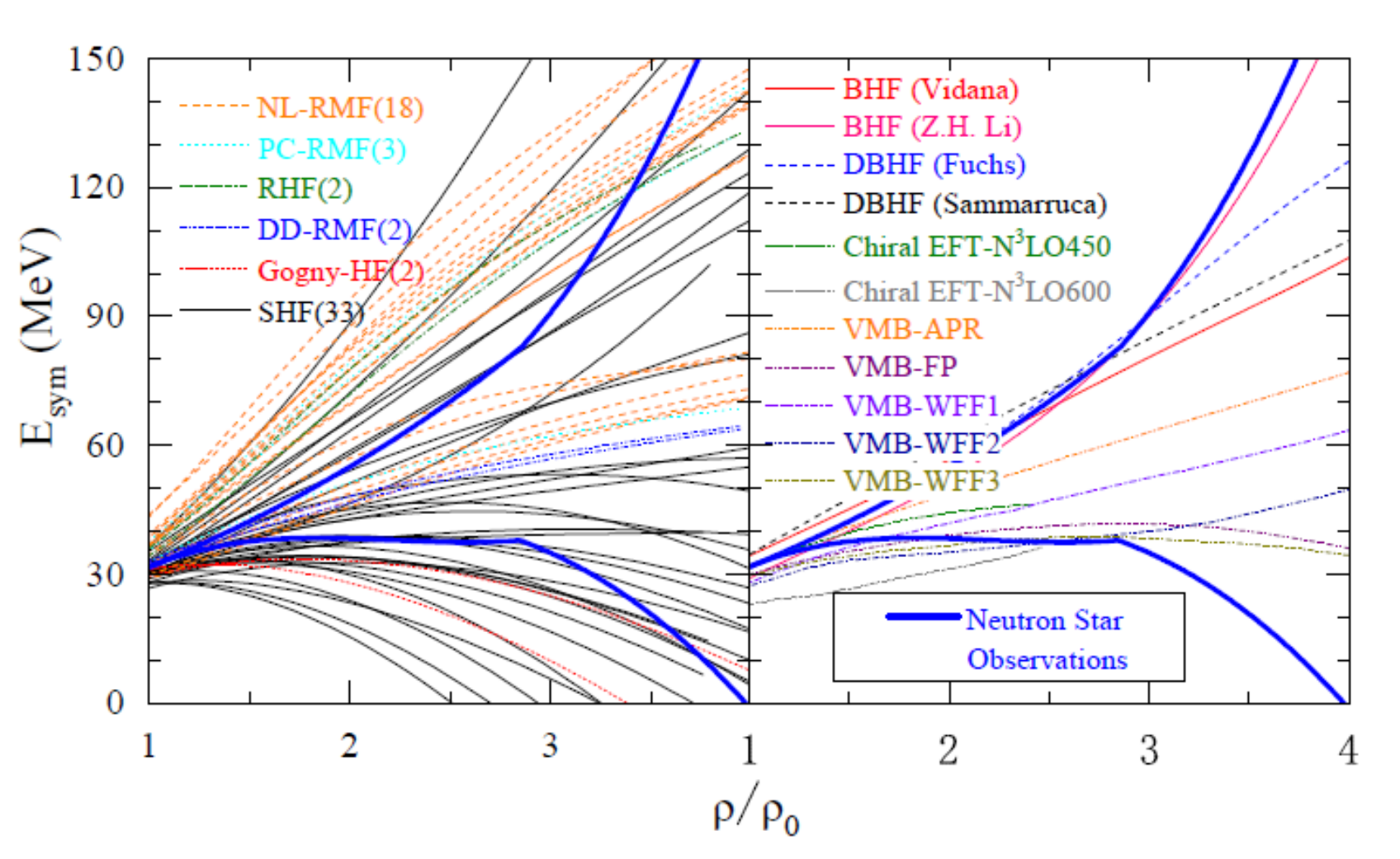} 
\includegraphics[width=7.4cm]{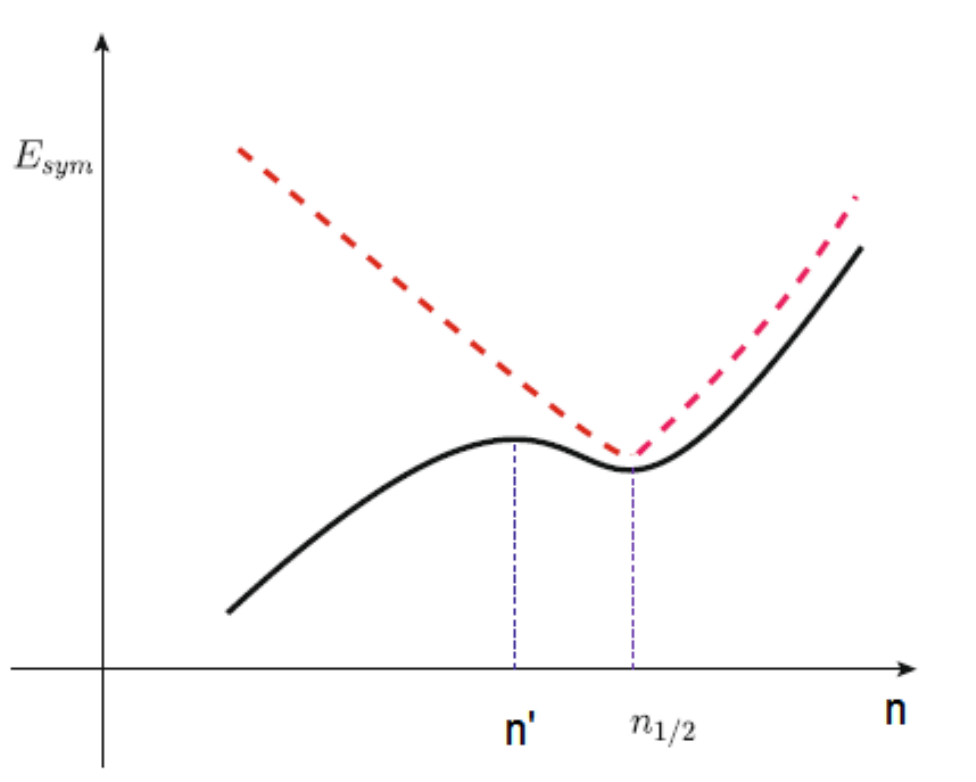}
\caption{Panorama of the symmetry energy $\Esym$. Left panel (copied from \cite{BAL-2021}): Wilderness in both various nuclear models and S$\chi$EFTs and bounds given by neutron star (up-to-date) observations (solid blue lines). Right panel: Schematic form of the cusp (dotted line) in skyrmion crystal~\cite{LPR-cusp}.  The solid line caricatures the effect of smearing by heavy degrees of freedom (HDsF). The interval between $n^\prime\gsim n_0$ and $n_{1/2}$ is the density regime that is arguably the most difficult to access by standard $\chi$EFT from below and by QCD proper from above as discussed in the text in connection with the sound speed and the tidal deformability.}
\label{Esym}
\end{figure*}

In the skyrmion crystal formalism, the topology change is associated with the behavior of the quark condensate at a density labeled $n_{1/2}$ which should, and generically does,  lie above $n_0$.  The quark condensate $\Sigma\equiv \la\bar{q}q\ra$,   nonzero both globally and locally for  $n <n_{1/2}$,  goes to zero at $n_{1/2}$ when space averaged,  $\phi_0\equiv \bar{\Sigma}\to 0$. But it is locally non-zero, thus generating chiral density wave and giving rise  to a non-vanishing pion decay constant, $f_\pi\neq 0$. This transition triggers  a  skyrmion in the matter to fractionalize into 2 half-skyrmions. Since the order parameter, here the pion decay constant,  is non-zero in the changeover, there is no low-order phase transition. This half-skyrmion ``phase"\footnote{This is a misnomer. Lacking a better terminology, however, we will continue to (mis)use this term.} resembles what is referred to as ``pseudo-gap phase" in condensed matter physics, e.g., in high-T superconductivity.  

An important --  and most crucial --  property of the symmetry energy in this formulation, $\propto 1/\lambda_{I}$, is that it develops a cusp at $O(1/N_c)$ at the density $n_{1/2}$ where $\phi_0\to 0$. The cusp structure seen in the skyrmion lattice simulation~\cite{LPR-cusp} is schematically depicted by dotted line in Fig.~\ref{Esym} (right panel). The exact location of the cusp depends on the parameters of the Lagrangian which are \`a priori unknown, so it is arbitrary. It will be determined later from neutron-star observations to lie within the range  $2 \lsim n_{1/2}/n_0 < 4$. This cusp form comes from an  interplay involving the behavior of $\phi_0$ between the quadratic derivative current algebra term and the countering contribution from the Skyrme quartic derivative term.  Roughly what happens is that  the increase of $\lambda$ from  the quadratic term as $\phi_0$  goes to zero is stopped by the quartic term at $n_{1/2}$ and starts dropping, causing the cusp in $1/\lambda$. It will be shown that this picture will be modified in nature by, among others, two observations. First, the skyrmion crystal simulation cannot be trusted at low density  below $n_{1/2}$, and next, the Skyrme quartic term can be taken as what results from integrating out HDsF from the skyrmion Lagrangian. This large $N_c$ consideration gives a remarkably simple $E_{\rm sym}$. 

What's noteworthy here is that it involves a standard nuclear theory reasoning.  We repeat here the argument because although It has been given in various publications by us (with other authors), it seems to remain un-understood. 

To illustrate what is captured in this cusp, we start by quoting  in Fig.~\ref{Esym} (left panel) the recent illuminating summary by B.A. Li et al.~\cite{BAL-2021} of the up-to-date experimental and theoretical status of the symmetry energy. It presents a giant wilderness.  All the theoretical models available up to date, e.g.,  various energy density functionals, $\chi$EFTs etc,,  fit  $\Esym$ (by fiat) to what's given in nature at $\sim n_0$. There are ample parameters available to allow it. The swamp sets in beyond $n_0$. Given the absence of trustful models --  not to mention theories, there is no guidance how the $\Esym$ should move at higher densities. There is nothing to prevent it from going up or down, even plunging below zero.  The current experimental observations such as neutron stars (and heavy-ion data limited to only a few times $n_0$) do not fare any better as indicated by the solid (blue) lines in the left panel of Fig.~\ref{Esym}. 

What is certain is that the cusp is buried in this jungle. It may not be absurd to think that the cusp structure could just be  an artifact of the lattice simulation. But it turns out, we will see, that it is not. When the jungle is cleared up by the  symmetries assumed to be involved, the cusp yields an extremely simple and portent mechanism needed for the EoS for massive stars. In particular, we will argue,  the cusp represents the hadron-quark ``duality" expressed in topology change~\cite{MR-review}.  The most striking consequence is that it will lead to two predictions, both neither confirmed nor falsified yet: One, what will be termed ``pseudo-conformal sound speed" and the other , baryon-charge fractionalized ``confined" fermions in the core of neutron stars.

We should stress that what's involved in our approach is  ``hadron-quark duality,"  not just hadron-quark continuity that captures crossover from hadronic degrees of freedom to quark/gluon degrees of freedom. In fact the notion of hadron-quark duality is a lot more general in the sense elaborated recently in the Cheshire-Cat Principle~\cite{CCP,dichotomy}. It represents the {\it necessity} at densities exceeding $n_0$  of the ``heavy degrees of freedom (HDsF)."    Those HDsF are to encode the quarks/gluons degrees of QCD at some high density without explicit presence of quark/gluons. How to do this precisely is presently unknown in the density regimes relevant to compact stars. This is because the densities involved are too far from the asymptotic regime where perturbative QCD is applicable and the only nonperturbative tool known, lattice QCD, is famously inaccessible at high density. So the question is: How does one proceed?
\subsection{Cusp Induced by Nuclear Tensor Force}
%
In \cite{LPR-cusp}, the cusp was reproduced by the role played by the pions and the vector mesons in the nuclear tensor force in standard nuclear structure physics. There the vector mesons were identified as hidden local fields and the scalar $\sigma$ as a dilaton, the Goldstone boson of spontaneously broken scale symmetry. The key idea there was to exploit the vacuum-change-induced density dependence in the sHLS Lagrangian in the presence of baryonic matter~\cite{BR91}\footnote {It seems highly appropriate to point out  here that the scaling relation proposed in this 1991 paper is still largely misinterpreted and misquoted in the literature. The indispensable role of hidden local symmetry locked to dilaton scale symmetry -- which was the key ingredient of the scaling-mass relation in medium, both in temperature and density -- was totally missed in applications of the scaling relation to nuclear processes under extreme conditions.  Up to date, the behavior of the vector meson ($V=(\rho, \omega)$) mass at high temperature as in dilepton processes in relativistic heavy-ion collision and at high density $n\gsim n_{1/2}$  as in CBM processes and compact stars was incorrectly treated, which led to the claim that the scaling relation was ``ruled out." The intricate property of hidden local symmetry dual to the gluons and the genuine dilaton nature of hidden scale symmetry, both of which figure crucially in this paper (and in \cite{MR-review}), is further elucidated, though not entirely vindicated, in the recent novel developments discussed in \cite{unified}.}.  Here we repeat essentially the same arguments to bring out certain characteristics of hidden sHLS  in the discussions, namely, the ``duality" assumed to hold \`a la Seiberg between hidden local gauge fields  and QCD gluons and a hadrons-quarks/gluons duality.  The objective is to link it to what S$\chi$EFT does at low energy (and density) and to extend it to higher densities where S$\chi$EFT is to break down. This would make our approach to compact-star matter in line with the spirit of the Folk Theorem on EFT. At present,  the duality assumed  is only a conjecture, but there are several compelling indications that such duality does most likely hold at high density (and perhaps also at high temperature)~\cite{komargodski, karasik,Y,kitano-matsudo}. In the absence of  a rigorous proof, we take this as our working assumption. 

%
\begin{figure*}[htb]
\centering
\includegraphics[width=7.5cm]{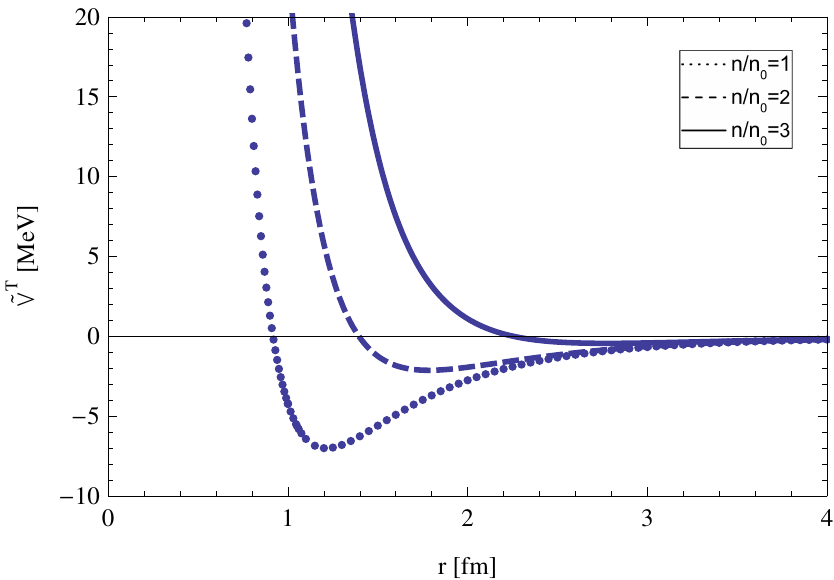}
\includegraphics[width=7.5cm]{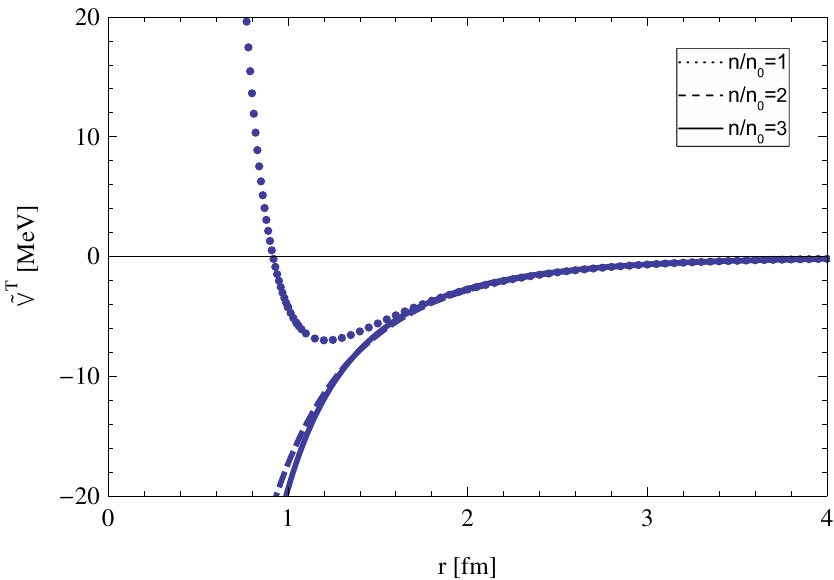}
\caption{Tensor force vs. density $n=(1-3) n_0$: Without topology change (left panel) and with topology change at $n_{1/2}\approx 2n_0$ (right panel)}
\label{tensorforce}
\end{figure*}
Our reasoning relies on two well-known (established) facts in nuclear physics in the presence of the HDsF. The first is that the symmetry energy is predominantly controlled by the nuclear tensor force,  and the second is that the nuclear tensor force gets principal contributions from the exchange of the pseudo-Nambu-Goldstone pion $\pi$ and the $\rho$ meson and coming  with an opposite sign, they tend to destructively interfere. It has also been established, within the framework of $Gn$EFT with density-scaling hadron masses~\cite{BR91}, that the net tensor force is to decrease with increasing density in the {\it effective range of force in nuclear medium with short-range correlations suitably taken into account}. What figures here are the ``vector manifestation" of the $\rho$ meson at high density encoded in HLS, the dilaton condensate controlling the hadron masses in dense medium and the interplay of the $\omega$-nucleon coupling with the nucleon mass~\cite{MR-review}. 

The resulting tensor force  is depicted in Fig.~\ref{tensorforce}. The left panel shows the decreasing tensor force at increasing density in the absence of topology change.\footnote{This dropping of the tensor force at increasing density is manifested in various nuclear phenomena, the validity of which has been amply supported.  A most spectacular case is the simple and elegant explanation of the long lifetime of C-14 beta-decay~\cite{holt}. There are some {\it ab initio} calculations using three-body forces that seem to explain more or less equally well, but this should not be considered belittling the beauty of the simple tensor-force mechanism. Correctly done, both are equally correct in physics.}. 
(We note for later discussion that the net tensor force would vanish in the relevant range at $n\sim 3n_0$). However if there intervenes the topology change at $n_{1/2}$, the tensor force undergoes a dramatic change as  seen Fig.~\ref{tensorforce} (right panel). For $n\gsim n_{1/2}$, the $\rho$ tensor gets suppressed more or less completely so that the net tensor gets abruptly recovered to that of the pionic strength.
How this changeover comes about is quite involved requiring a series of arguments ~\cite{MR-review}, but it is not hard to understand what'a at work in the mechanism with two  assumptions. The assumptions are that (A) the vector mesons introduced as HDsF are hidden local symmetric subject to ``composite gauge symmetry"~\cite{HY:PR,suzuki}  and (B) the scalar that provides an attractive nuclear force is the dilaton $\sigma$ of the ``genuine dilaton (GD)"~\cite{GD} (or ``conformal dilaton (CD)"~\cite{DDZ}.\footnote{The ``conformal dilaton"  is proposed to be in a phase embedded in the conformal window having an IR fixed point of the same type as that of GD. What we are concerned with here is an ``emergent" scale symmetry coming from strong nuclear correlations. We speculate that in dense matter, the putative CD could merge with the GD. More on this in the Conclusion section.})
Now the assumption (A) asserts that at some high density, the vector mesons become massless,  in particular  with the gauge coupling $g_\rho$ going to zero~\cite{HY:PR} associated with the vector manifestation fixed point mentioned above,  and the assumption (B) admits a (precocious) emergence of spontaneously broken scale symmetry with $f_\sigma\approx f_\pi\neq 0$, accommodating  massive matter fields, in particular, light-quark baryons, \`a la genuine dilaton scenario with the dilaton condensate dictating how hadron masses scale in density~\cite{BR91}. The two effects entail the abrupt changeover at $n_{1/2}$ in the tensor force in Fig.~\ref{tensorforce} from the left panel to the right panel.

To see how this changeover produces the cusp in $\Esym$, one recalls that the symmetry energy is predominantly controlled by the tensor force. A quick and simple way to estimate the dominant tensor-force contribution to $\Esym$  is to do the closure-sum approximation of the iterated tensor force terms~\cite{closuresum}. This exploits that the ground state is strongly coupled by the tensor force (subject, however, to the decreasing strength with density described above) to the particle-hole states of excitation energies $\sim 200$ MeV, so 
\be
E_{\rm sym}\approx C\frac{\la V_T^2\ra}{200\ {\rm MeV}}
\ee
with $C> 0$ a known constant.  With the NN interactions duly screened by short-range correlations (for which the $\omega$ meson enters) via RG, it can be seen that $\la V_T^2\ra$ decreases as density goes toward $n_{1/2}$ and then increases afterwards in the precise way as in the skyrmion lattice simulation, thus reproducing the cusp Fig.~\ref {Esym} at $n_{1/2}$. While this argument holds more reliably  on the right side of the cusp, namely in the half-skrymion phase, it  is not expected to to hold well in the skyrmion phase away from the cusp. This is because there the effects well described by S$\chi$EFT that include complicated configurations at high chiral orders involving other components of the force than the tensor-force are missing in this treatment. This will become visible  in the $Gn$EFT result  shown below.
\subsection{Smoothed Cusp}
\begin{figure*}[t]
\centering
\includegraphics[height=6.2cm]{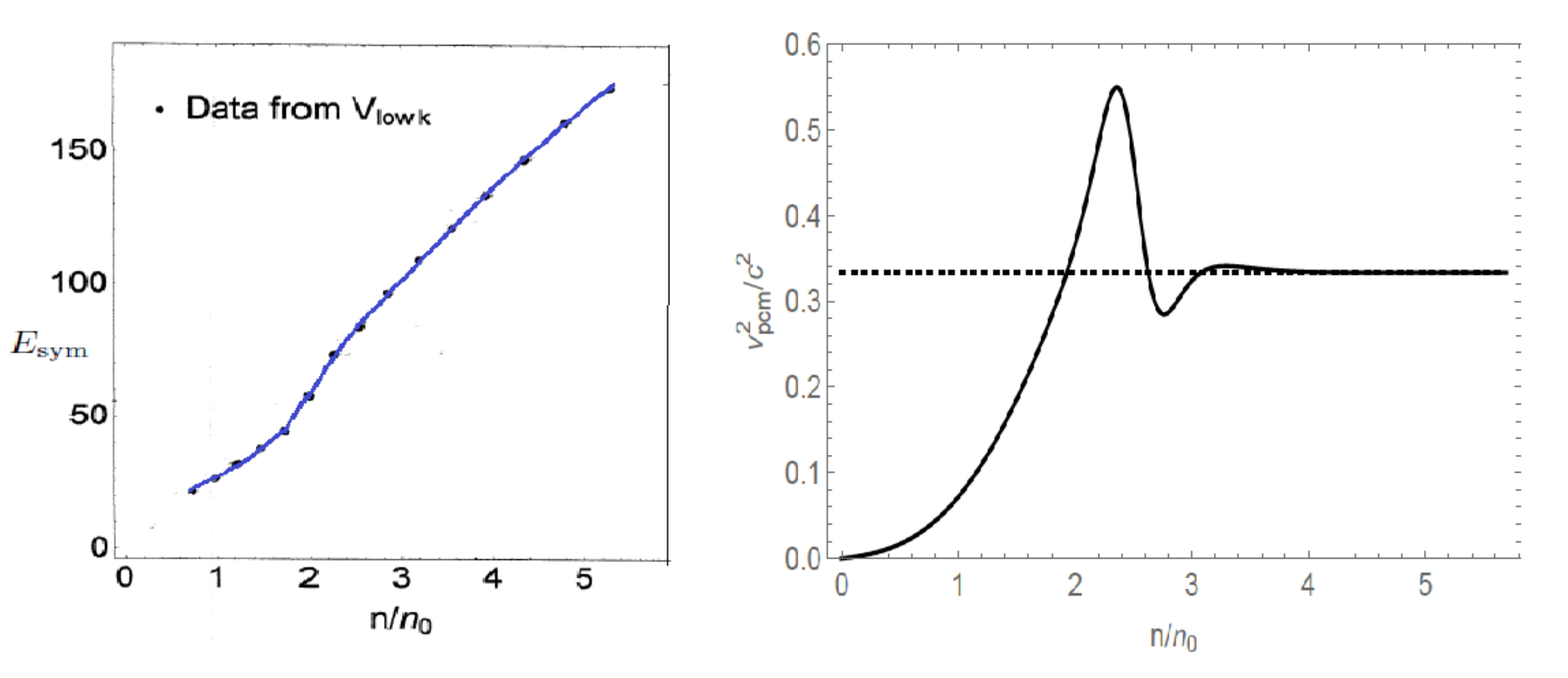}
\caption{ $\mathbf{E_{\rm\bf sym}}$ (in MeV)  and $v_{pcs}^2/c^2$  for neutron matter ($\alpha=1$) calculated in $Gn$EFT for $n_{1/2}\sim (2.0- 2.5) n_0$. The bump samples the density range between $n^\prime$ and $n_{1/2}$ in Fig.~\ref{Esym} (right panel). Note that the structure of $v_{pcs}$ follows directly and entirely from  that of $E_{\rm sym}$  for $n\gsim n_{1/2}$ as explained in the text.}
\label{EsymGnEFT}
\end{figure*}

This calculation for the cusp (with the nuclear tensor force affected by the topology change) smoothed by the HDsF corresponds to the large $N_c$ and quasi-classical approximation in standard nuclear physics calculations. In the formulation of $Gn$EFT, this is equivalent to the mean-field approximation with the sHLS Lagrangian~\cite{MR-review} which corresponds to the Landau Fermi-liquid fixed point approximation in the large $N_c$ and large $\bar{N}\equiv k_F/(\Lambda-k_F)$ limit~\cite{shankar}.  
As shown in Fig.~\ref{EsymGnEFT} (left panel) this cusp is made to smoothly cross over in the ``$V_{lowk}$ renormalized group (RG) approach" going beyond the Fermi-liquid fixed point approximation in $Gn$EFT employed in \cite{PKLMR,MR-review}. It takes into account $1/\bar{N}$ corrections  in  the ``ring-diagram" approximation. It corresponds to a generalized Fermi-liquid theory applicable to the relevant range of densities from $n_0$ to the compact-star matter density $\sim (5-7)n_0$ with the topology change incorporated at $n_{1/2}$. It is strictly valid in the large $N_c$ limit but has been verified to work well for nuclear matter, arguably as well as the S$\chi$EFT to N$^3$LO. The power of this approach is that while  the S$\chi$EFT most likely breaks down at $n_{1/2}$,  the $Gn$EFT approach becomes more reliable at higher densities as the Fermi-liquid fixed point is approached, that is as $1/\bar{N}\to 0$.\footnote{ The role sHLS plays here in smoothing the cusp is albeit inexplicably analogous to eliminating the  cusp singularity in the $\eta^\prime$ potential term for the $\eta^\prime$ EFT with the HLS fields becoming topological Chern-Simons fields, giving rise to the fractional quantum Hall droplet baryon~\cite{karasik,kitano-matsudo}.} Although the cusp is smoothed, it makes the symmetry energy that is soft below $n_{1/2}$ to stiffen above $n_{1/2}$. This not only accounts for the observed massive neutron stars but as we will show, will {\it render moot} the nLORE, hence resolving the  PREX-II dilemma.  What's even more striking is that the cusp impacts via the $\Esym$ so constructed the sound speed $v_{pcs}$ as
\be
 E_{\rm sym}(n)\to v^2_{pcs} (n)/c^2\sim 1/3\ {\rm for }\ n\gsim n_{1/2}.
\ee
This is because $E_{sym} (n)=E(n,\alpha=1)-E(n,\alpha=0)$ from Eq.~(\ref{eq.Esym}) and the trace of the energy-momentum tensor (TEMT) given by $E(n,\alpha)$ is density-independent for $\alpha=0$ and $\alpha=1$ for $n\gsim n_{1/2}$~\cite{MR-review}.  Hence the crossover in the latter directly impacts the bump in $v_{pcs}$. These matters will be taken up in Sect.~\ref{bump}.

\section{How $Gn$EFT fares}\
In order to give credence to the $\Esym$ obtained in $Gn$EFT that we will rely on,  we summarize what the $Gn$EFT treated  in $V_{lowk}$RG predicts for the EoS for nuclear matter and how it fares in nature. As stressed in \cite{MR-review},  the possible topology change density is constrained to the range $2\lsim n_{1/2}/n_0 < 4$. For simplicity we take $n_{1/2}\gsim 2 n_0$  as representing our prediction within  a small range of uncertainty.

In what follows, the strangeness flavor degrees of freedom,    hyperons as well as kaons,    do not enter in the density regime involved.  The reason for this is explained in  Conclusion Section.

We divide the density regime into two: (A) $n\gsim n_0$ and (B) $n\gsim n_{1/2}=2n_0$.
\begin{itemize}
\item (A)  Up to the topology change density $n_{1/2}\gsim 2 n_0$, there is only one parameter which is completely determined by how the pion decay constant $f_\pi$ scales with density; it  is known up to $n\sim n_0$.  Within a bit of fine-tuning on this scale parameter, all the EoS properties come out fully consistent with the accepted values : They are $n_0=0.16$ fm$^{-3}$, BE =16.7 MeV, $K_0=250 (240\pm 20)$ MeV,  $J=\Esym (n_0)=30.2 (31.7\pm 3.2)$ MeV, $L=67.8 (58.7\pm 28.1)$ MeV.
Given in the parenthesis are quoted -- for the illustrative purpose -- from the recent compilation by Zhang and Li~\cite{BAL}. The same analysis gives the comparison at $n=2n_0$:  $\Esym (2n_0)= 56.4 (50.55\pm 5.99)$ MeV.  This will be relevant for our argument given below.  
\item (B) For $n > n_{1/2}\approx 2 n_0$, there are effectively two additional scaling parameters, one for the coupling constant $g_A$ and the other for the $\omega$-meson gauge coupling which differs from the $\rho$ gauge coupling that flows to  the vector manifestation fixed point $g_\rho=0$. Both are intricately correlated with the emergent scale symmetry~\cite{MR-review}. This does not affect what follows below, so we won't go into details here. 

The predicted star properties are\footnote{There is a possible caveat in what is quoted as our predictions for the relation between the radii $R$ and masses $M$, particularly for GW data. It is argued~\cite{crust} that to make a quantitatively reliable analysis on $R$ vs. $M$, the EoSs of the core and crust should be treated thermodynamically consistently. This consistency has not been imposed in \cite{PKLMR} from which we are quoting the predicted values where the crust-core transition was taken at $n_{\rm core-crust}\approx 0.5 n_0$.  This caveat might be relevant to the quantities $\Lambda_{1.4}$ and $R_{1.4}$ but most likely less  for other macrophysical quantities of massive stars.}: Maximum star mass $2.05 \lsim M_{max}/M_\odot \lsim 2.23$ for $2.0\lsim n_{1/2}/n_0 < 4.0$,  radius $R_{2.05M}\approx 12.0$ km,   $\Lambda_{1.4}\lsim 650$, $R_{1.4 }\approx 12.8 $ km. 
\end{itemize}

%

\section{Heavy degrees of freedom as dual to gluons and quarks:  Hadron-quark continuity}
The issue of possible resolution to the PREX-II dilemma in our approach which is closely linked to also other issues currently in discussion in the literature is encapsulated in $\Esym$ in Fig.~\ref{EsymGnEFT} (left panel).  It is given by the generalized $Gn$EFT that involves {\it only one} Lagrangian with the HDsF suitably incorporated together with the topology change. It contains no phase transitions in the sense of Ginzburg-Landau-Wilsonian paradigm, but it is taken to simulate hadron-quark/gluon continuity. Here we are quoting the result obtained for the crossover density $n_{1/2} \sim (2.0- 2.5) n_0$.  For the semi-quantitative aspect we are addressing here,    the conclusion we arrive at is essentially the same for the range  $2\lsim n_{1/2}/n_0 <4$. 
\subsection{Heavy Degrees of Freedom\\ and Correlated Fermions}
\begin{figure*}
    \centering
    \includegraphics[width=0.4\textwidth]{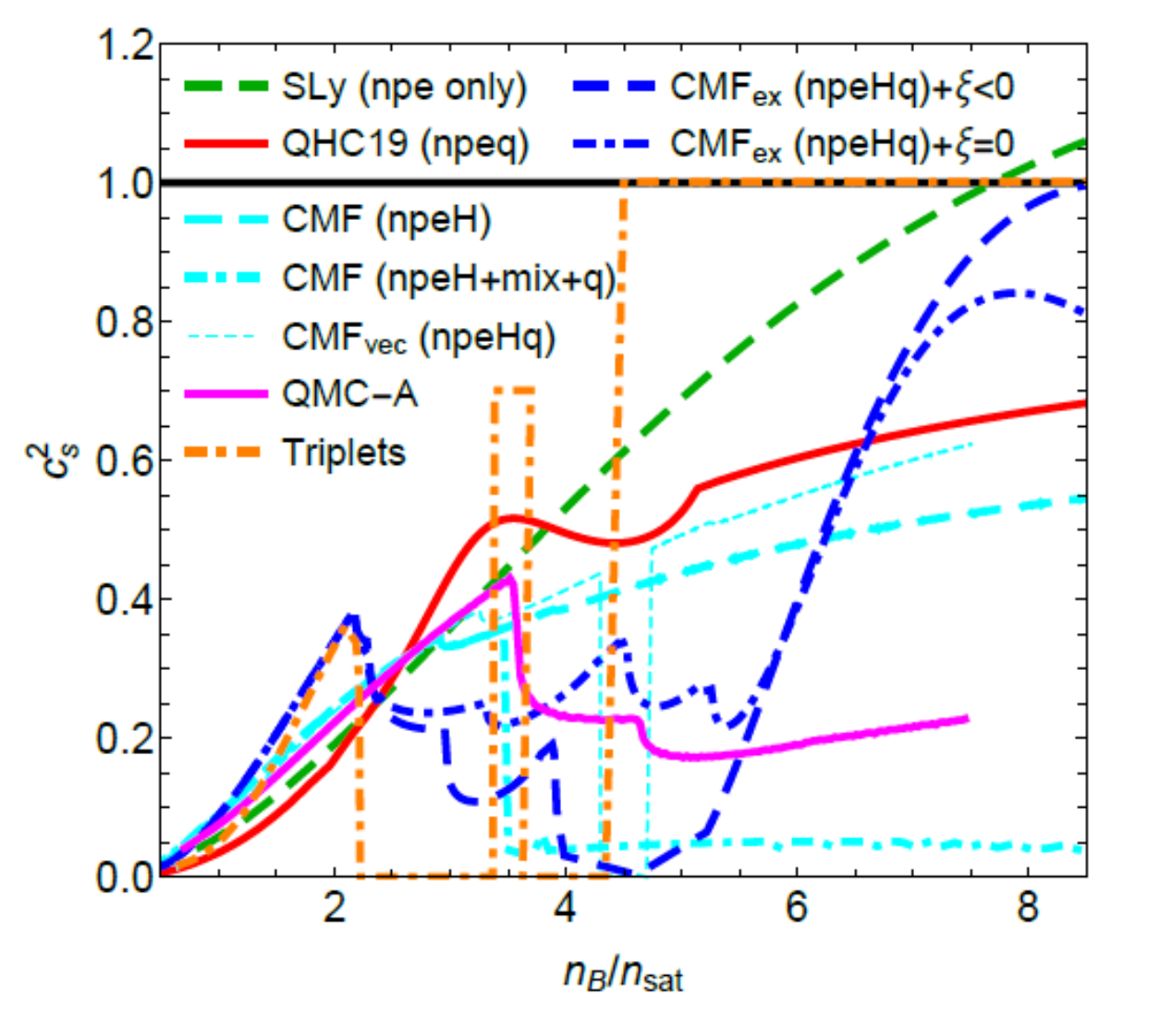}
    \includegraphics[width=0.52\textwidth]{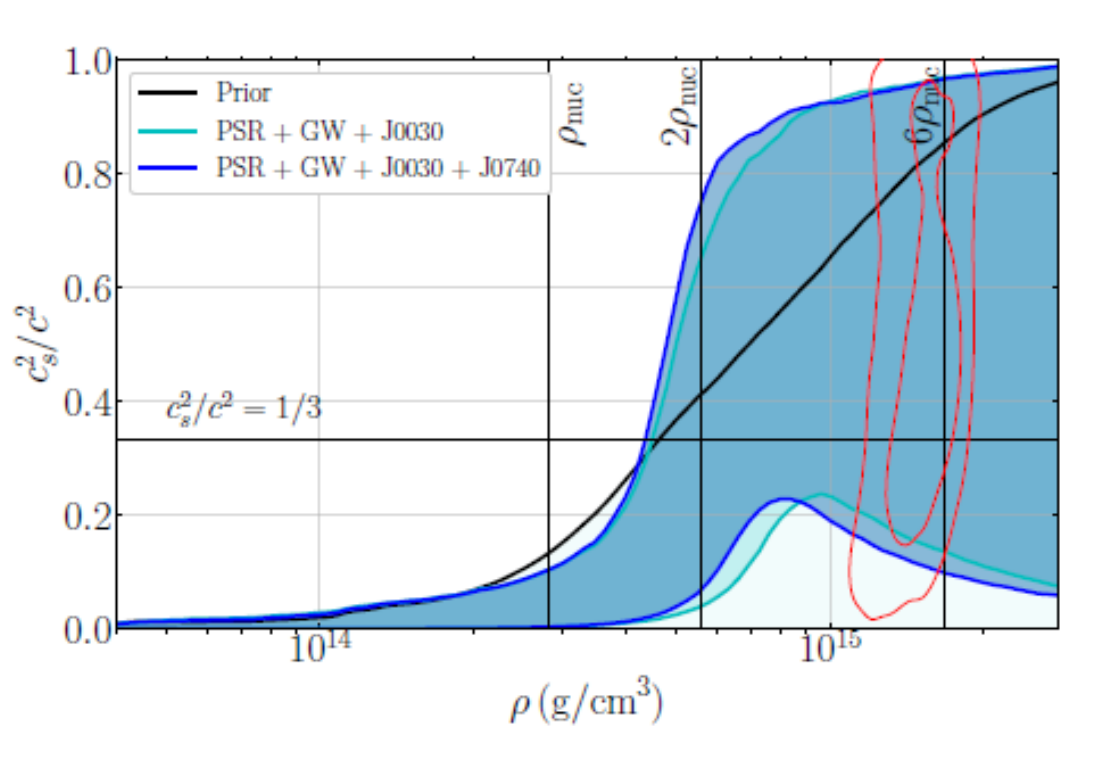}
    \caption{Plethora of bumps, spikes, skates and what not in the sound speed $c_s\equiv v_s$:   (left panel) Wilderness for massive star with $M > 2.5 M_\odot$ with and without phase transitions assuming it is  a stable neutron star  instead of a black hole (copied from \cite{bump-orgy}); (right panel) strongly interacting baryonic matter in the core (copied from \cite{bump-impact}) where $\rho$ is $n$ in unit of $g/cm^3$. }
\label{bumps}
\end{figure*}
In addressing the issues involved, there are two important points to note: 

First, the heavy-degrees of freedom smoothen (or do away with) the cusp ``singularity"  with the vector mesons playing the (dual) role of the gluons and induce the crossover from soft-to-hard in the EoS at the transition region.  As mentioned, the maximum that can be reached in $Gn$EFT is $M_{\rm max}\approx 2.23 M_\odot$.  At this crossover density, however, the maximum of the bump/spike  in the sound speed exceeds the causality bound at $n\sim 3.5 n_0$, so may not  be  physically acceptable although no other star properties seem to go haywire. This implies that our approach will get into tension with $\gsim 2.5 M_\odot$ stars should they be confirmed to be stable neutron stars.

Second, totally distinctive from all currently available ones in the literature, the present EoS unambiguously predicts~\cite{MR-review} what we call ``pseudo-conformal sound speed" $v_{pcs}^2/c^2\approx 1/3$ for density $n\gsim n_{1/2}$ depicted in Fig.~\ref{EsymGnEFT} (right lane).  It is the solid line in Fig.~\ref{EsymGnEFT} (left lane) that connects the numerically obtained $V_{lowk}$RG ``data" that precisely gives the sound speed $v_{pcs}^2/c^2$ that converges to 1/3 at $n\sim 3n_0\gsim n_{1/2}$ and stays at 1/3 beyond the central density $\sim 5n_0$ of the star.  This is an unequivocal impact of the symmetry energy on the pseudo-conformality of the sound speed.

It should be noted that $v_{pcs}^2/c^2=1/3$ here does {\it  not} represent the conformal sound speed associated with the vanishing trace of the energy-momentum tensor (TEMT). It is not the ``conformal sound-speed bound" that is referred to in the literature in addressing the role of ``deconfined quarks" in the core of dense neutron stars.  In the system we are dealing with here, the TEMT cannot go to zero in the density regime involved since it is still far from the (putative) IR fixed point~\cite{GD}.  We underline here that $v_{pcs}^2/c^2=1/3$ reflects pseudo-conformality emergent from {\it strong correlations} involving the degrees of freedom including the HDsF that give rise to nearly non-interacting quasi-fermions~\cite{MR-review,MR-review}.  It embodies hidden scale symmetry.  It is far from the state of nearly free ``deconfined" quarks discussed in the literature where first-order phase transitions are invoked.  It depicts a strongly correlated matter in a way resembling what takes place in certain condensed matter physics.

\subsection{Bumps of Sound Speed}\label{bump}
There are a great deal of discussions currently in the literature on the impact on the EoS of dense baryonic matter in terms of the structure of the sound speed in the crossover region  in density $\gsim 2 n_0$. From the point of view of nuclear physics, the problem here, as mentioned above, is that from the crossover region indicated  between $n^\prime$ and $\gsim n_{1/2}$ in Fig.~\ref{Esym} for hadrons-to-quarks/gluons to the core density of massive stars is the density regime  which is the hardest to access theoretically, both bottom-up in EFT and top-down in QCD. Chiral effective field theory S$\chi$EFT works reliably for nuclear matter properties, but it is very likely to break down at this crossover region.  Top-down, the perturbative QCD must also break down at the star-matter density and will certainly be inapplicable at the crossover region.  Thus it is not totally absurd to come up with various wild  scenarios  in the region involved.  

In fact, much discussed are the plethora of bumps, spikes, kinks etc. of various sizes ranging from the crossover to the core density region of star with or without phase transitions. 

Among many others found in the literature, we pick two cases  for illustration in Fig.~\ref{bumps}. 

The left panel shows the possibilities of massive stars $M > 2.5 M_\odot$ with bumps of the sound speed all the way from zero to violating the causality bound, typically involving phase changes~\cite{bump-orgy}. Our approach, as mentioned above, cannot access this mass star within the framework we are working with.  It will require a major revamping to accommodate such massive stars if they exist.

On the contrary, the right panel illustrates the case without phase changes (or with smooth crossover) that displays the sound speed largely violating the conformal bound $v_s^2/c^2=1/3$ starting from the crossover region~\cite{bump-impact}.  Since the issue of the PREX-II is related to what's treated in \cite{bump-impact}, this case is highly relevant.  The analysis of \cite{bump-impact} relies on what is called ``non-parametric model based on Gaussian processes,   un-tied to specific nuclear models, not subject to systematic errors and possesses wide-range intra-density correlations and targets wide-range of densities."  While it is not clear to us what this model means with respect to our  approach, there is a striking difference between the two.  It is in the structure of the constituents in the core of massive stars.

 For comparison with our prediction, we make a  list of some of the results reached by the analysis~\cite{bump-impact} on the most massive star known so far, i.e., J0740+6620 (NICER+XMM-Newton):
 $M_{max}=2.24^{+0.34}_{-1.06} M_\odot$, $R_{1.4} =12.54^{+1.01}_{-1.06}$ km,  
$ \Delta R=R_{2.0}-R_{1.4}=-0.04^{+0.81}_{-0.83}$ km  and 
$n_{\rm cent}=3.0^{+1.6}_{-1.6} n_0$. Based on these and other considerations, the authors of \cite{bump-impact}  arrive at the conclusion that the conformal sound speed bound is {\t strongly violated} as depicted  in Fig.~\ref{bumps} (right panel) reaching the maximum 
\be
v_s^2/c^2=0.79 ^{+0.21}_{-0.20}.
\ee
This strong deviation from the conformal sound speed is attributed by the authors to ``strongly interacting hadronic degrees of freedom" that the authors interpret as  ``disfavoring" the appearance of ``explicit" QCD degrees of freedom in the core of stars.  This property is consistent with the low central density $\sim 3 n_0$ found in the analysis.  One can consider the PREX-II dilemma (\ref{nonnormal}) to belong to this class of scenario.

It should be admitted that given the total paucity of theoretical tools applicable in that density regime, perhaps one cannot rule out this possibility.

However what $Gn$EFT has predicted is strongly and fundamentally different from what's found in \cite{bump-impact}. As summarized in \cite{MR-review}, our pseudo-conformal structure yields the following:  $M_{max}\approx 2.24 M_\odot$, $R_{1.4} \approx 12.8$ km,  $\Delta R=R_{2.0}-R_{1.4}\approx - 0.08$ km and  $n_{\rm cent}\approx 5.1 n_0$. Thus except for one quantity, the central density,  the overall macrophysical properties predicted are globally the same as those of \cite{bump-impact}. But the sound speed of the star (see Fig.~\ref{EsymGnEFT} (right panel) and Fig.~\ref{bumps} (right panel)) is {\it drastically} different. The closeness in the $Gn$EFT approach of the pseudo-conformal sound speed to the conformal speed bound together the higher central density  in contrast to what's expected of hadronic constituents~\cite{bump-impact} signals fractionalized quasi-fermions different from baryonic matter. The structure of the constituents of the star core resembles that of ``deconfined" quarks but the non-zero trace of the energy-momentum tensor makes it basically different from the conformal state.
\begin{figure*}
\centering
\includegraphics[width=5.6cm]{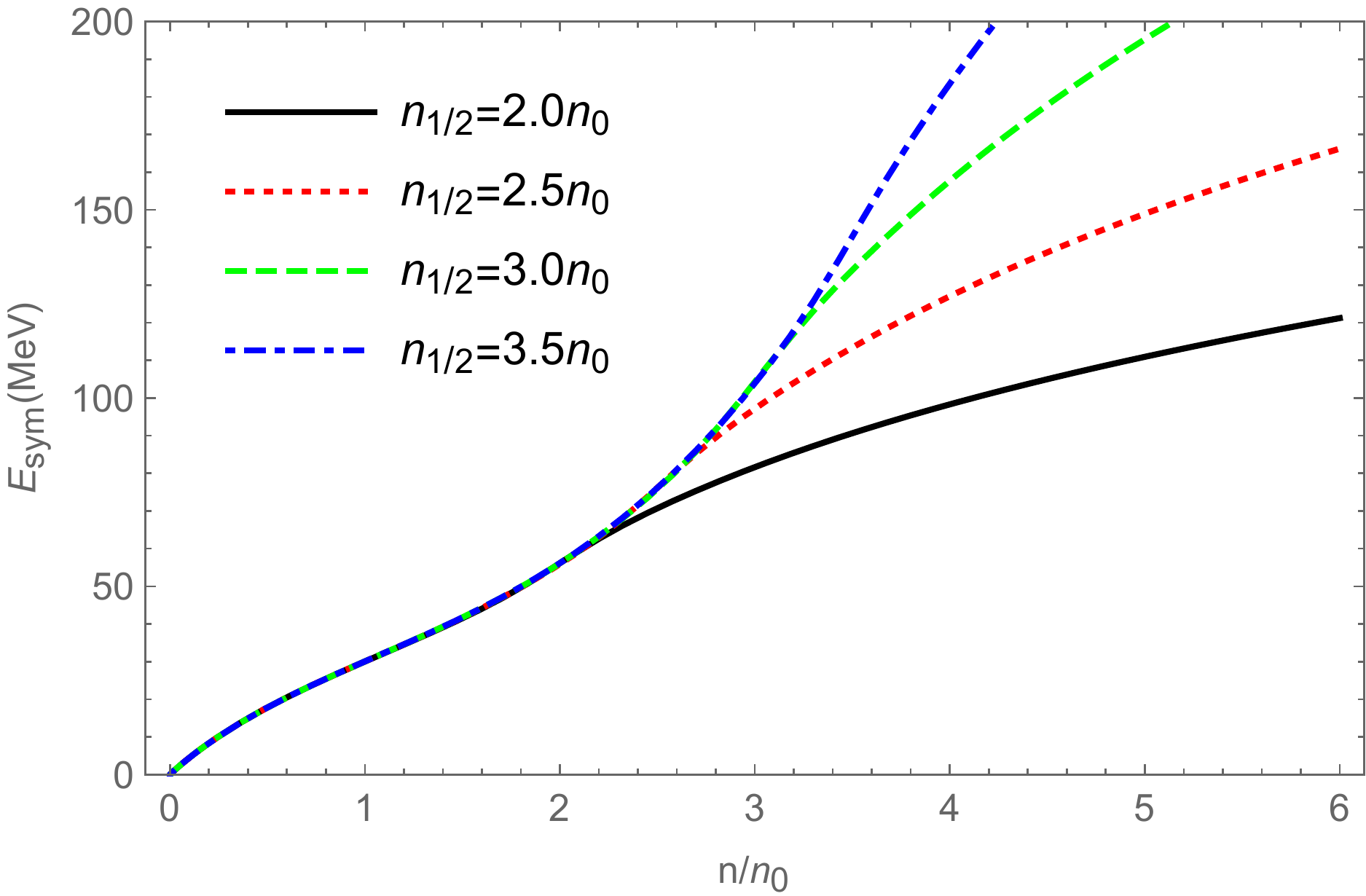}
\includegraphics[width=5.6cm]{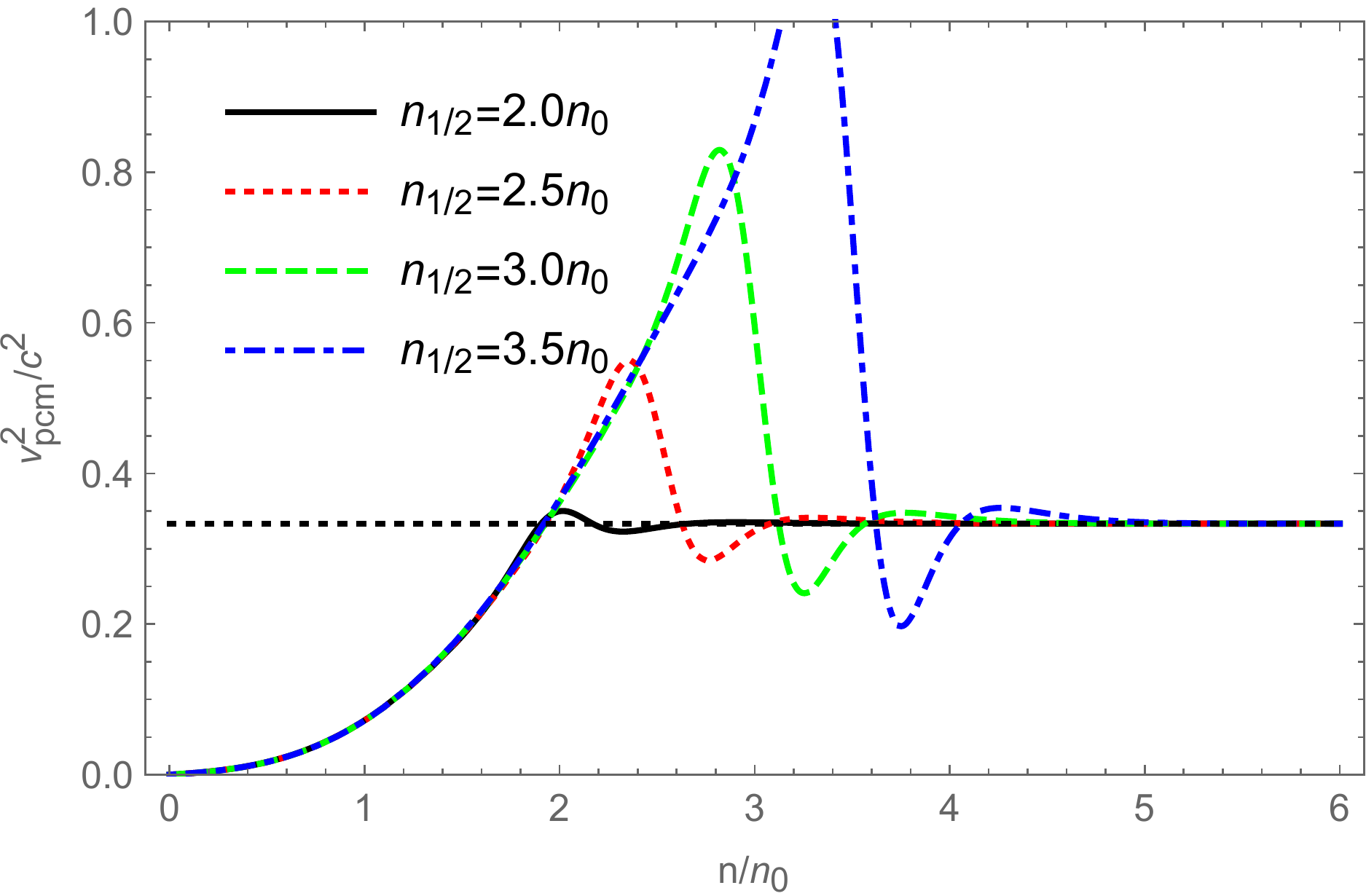}
\includegraphics[width=5.6cm]{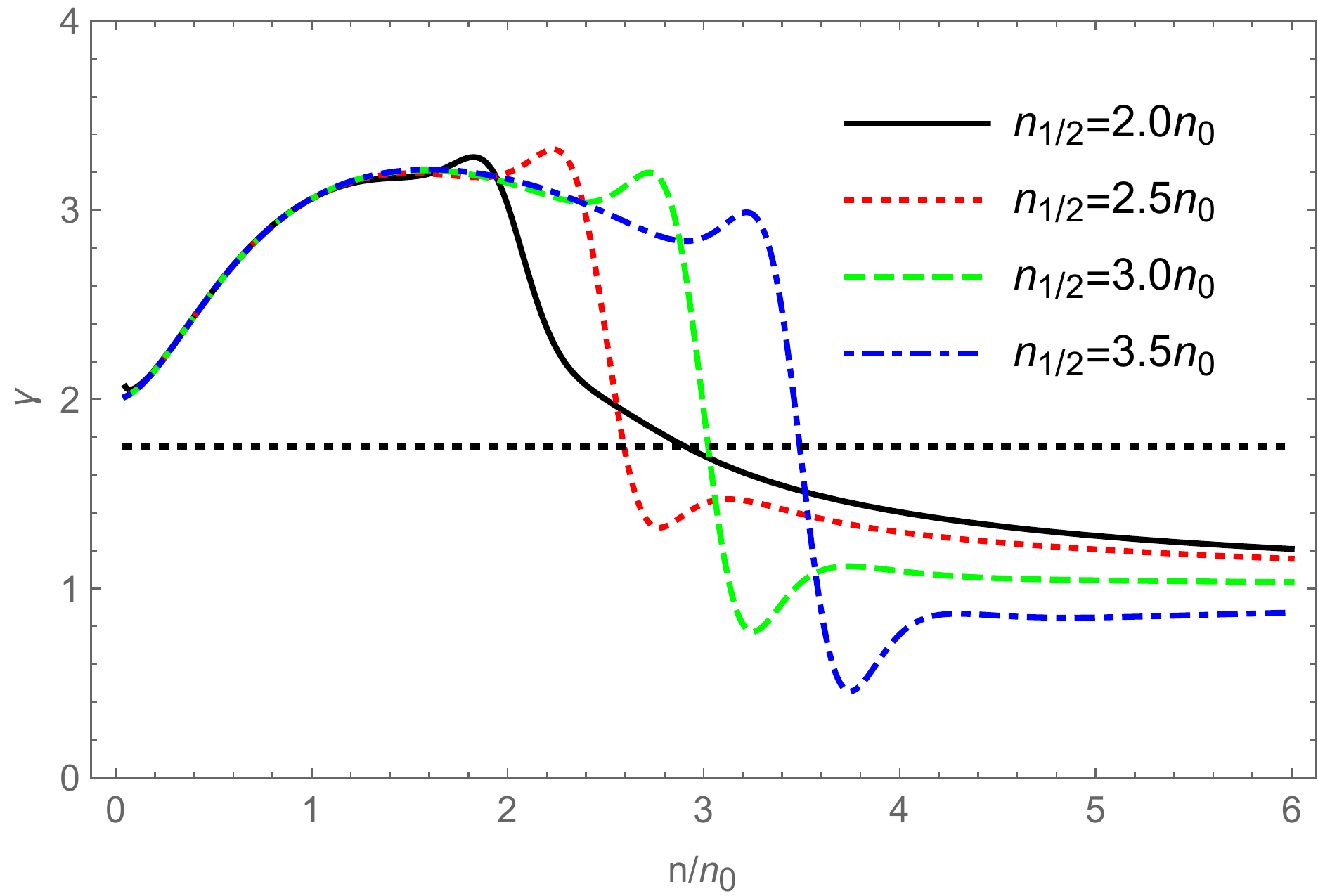}
\caption{ $E_{\rm sym}$  (left panel), sound velocity $v^2_{pcv}/c^2$ (middle panel) and polytropic index $\gamma$ (right panel) as function of density for $n_{1/2}/n_0= 2.0, 2.5, 3.0$ and 3.5.}
\label{EoSS}
\end{figure*}

An illuminating observation can be made when one looks at what happens with different topology-change densities. Taking the $V_{lowk}$RG predictions for baryonic matter for densities $\lsim n_{1/2}$, assuming the EoS for $n > n_{1/2}$ to be pseudo-conformal\footnote{This pseudo-conformal structure was predicted in the $V_{lowk}$RG formalism going beyond the large $\bar{N}$ limit (i.e., Fermi-liquid fixed point limit) for $n_{1/2}/n_0=2.0$ and 2.5~\cite{PKLMR}. The assumption here -- which we believe is reasonable -- is that the pseudo-conformality holds as well for $n_{1/2}/n_0 > 2.5$.},  one can readily compute $E_{\rm sym}$, the sound speed $v_s^2$ and the polytropic index $\gamma=d\ln P/d\ln \epsilon$ for various $n_{1/2}$.  As shown in  \cite{PKLMR},  the pseudo-conformal energy per baryon $E/A$ for $n\geq n_{1/2}$ can be parameterized as
\be
(E/A)|_{n \geq n_{1/2}}  = {} - m_N +  B\left(\frac{n}{n_0}\right)^{1/3} + D\left(\frac{n}{n_0}\right)^{-1}
\label{EII}
\ee
with the coefficients $B$ and $D$ fixed by continuity between $V_{lowk}$RG and  (\ref{EII}) at $n_{1/2}$. The results comparing $E_{sym}$, $v_{pcs}$ and $\gamma$ are plotted for the range of the topology change density $2.0\leq n_{1/2}/n_0\leq 3.5$ in Fig.~\ref{EoSS}.

Here are two observations to make here. 

One is that $E_{\rm sym}$ is insensitive  to the topology change density  $n_{1/2}\gsim 2 n_0$ up to $n\sim 2.5n_0$ but is very sensitive at higher density to the density $n_{1/2}$.  The greater $n_{1/2}$ the harder $E_{\rm sym}$ becomes.  Thus $J$ and $L$ quoted above are more or less independent of the cusp density. This point is  counter to the nLORE vis-\`a-vis the PREX-II dilemma. 

The second observation is that the sound speed and the polytropic index clearly show how the pseudo-conformality sets in for different $n_{1/2}$.  It is surprising that the causality bound is violated already at $n_{1/2}/n_0\sim 3.5$.  Furthermore the $E_{\rm sym}$ for $n_{1/2}/n_0\sim 3$ is most likely more repulsive than the bound indicated in the current analysis~\cite{BAL-2021} at densities $n\gsim 2n_0$.  Future experiments will either support or rule out this prediction. What seems striking is that the topology change density -- a.k.a. the hadron-quark continuity density -- is narrowed to a small window $2.0\lsim n_{1/2}/n_0 < 4$.

These observations 
\subsection{Topology Encodes Microscopic Dynamics}\label{topology-encoded}
The principal advantage of our approach\footnote{We believe that what we are discussing here would be little, if any, affected by the caveat associated with the crust.} is that it relies on what is likely a robust topological structure which provides a coarse-grained macroscopic description of what is presumably taking place in the density regime more or less uncontrolled by a microscopic theory\footnote{The crucial role of topology played here has an analogy in condensed matter physics. For instance in the fractional quantum Hall effects Chern-Simons topological field theory captures the microscopic structure of, say, Kohn-Sham density functional theory~\cite{mapping}.}. In our approach  the sound speed does produce the simple bumps  of  Fig.~\ref{EoSS} caused by the  intervention of the HDsF  dual to QCD degrees of freedom~\cite{MR-review} with its characteristics encoded  in $E_{\rm sym}$ capturing the crossover density $n_{1/2}$.  For the case of $n_{1/2}\simeq 2.0(3.0) n_0$, it is a bump reaching $v_s^2/c^2\sim 0.7 (0.8)$. But for $n_{1/2}\sim 3.5 n_0$, as mentioned, it is a spike with the maximum of which going out of the causality bound. Yet despite the different bump heights in the range $2\lsim n_{1/2}/n_0 < 4$ (even violating causality bound for the case $n_{1/2}~ 3.5 n_0$),  the sound speed $v_{pcs}^2/c^2$ converges in all cases to 1/3 slightly above $n_{1/2}$ with the global star properties not noticeably different between them.  The appealing aspect of our prediction~\cite{MR-review} is that it is an extremely economical description -- coarse-graining the  microscopic details of what's found in \cite{bump-orgy,bump-impact} -- that could be capturing the underlying physics.  We are arguing that it is precisely the {\it correlated strong interactions} leading to the Landau Fermi-liquid {\it quasiparticle} structure which becomes more valid  with increasing density after the topology change as $\bar{N}=k_F/(\Lambda_F-k_F)\to \infty $~\cite{shankar}, manifesting  pseudo-conformal symmetry in the sound speed.  At much higher densities approaching the putative IR fixed point, however, the Fermi-liquid structure should break down as in condensed matter physics~\cite{non-fermi}.

The question that can be raised here is how can the physics of the complexity in the sound speed  favored by \cite{bump-impact}  be
reproduced  by the extremely simple structure driven by the emerging hidden scale symmetry that leads to Fig.~\ref{EsymGnEFT} (right lane)? The possible answer to this could be that the macroscopic properties of massive neutron stars are in some way insensitive to the microscopic details of the bump structure of the sound speed with the emergent symmetries manifesting in the sound speed related to what's operative in the ``quenching of $g_A$" in baryonic matter mentioned below.  We come back to this issue in the Conclusion section.

Needless to say, as coarse-grained, there can be fluctuations on top of 1/3 coming from corrections to the underlying scale symmetry. That the sound speed converges precociously  to $v_{pcs}^2/c^2\simeq 1/3$  could be an oversimplification.  First of all the density involved $< 10 n_0$ is far from the density at which the vector manifestation limit and/or the dilaton limit fixed point is approached~\cite{MR-review}, so the scale symmetry should be broken (as indicated by the effective dilaton mass which must be substantial counterbalancing the $\omega$ repulsion in dense medium). However there is an indication that scale symmetry can be ``emergent," even if not intrinsic,  in certain highly correlated nuclear dynamics. One prominent evidence for it was seen in the so-called ``quenched $g_A$" phenomenon in nuclear beta decay~\cite{gA}. The effective $g_A$ in nuclear Gamow-Teller transition in light nuclei $g_A^\ast\approx 1$ arises due  to strong nuclear correlations influenced by hidden scale symmetry reflected in low-energy theorems.  The approach  $g_V\to g_A=1$ at high density $\gsim 25n_0$ as the dilaton limit fixed point is approached is closely correlated to how the quenched $g_A$ results in finite nuclei~\cite{multifarious}.  Furthermore  that the simple pseudo-conformal sound-speed structure ``governed" by the $\Esym (n)$  setting in slightly above the crossover density with none of the compact-star properties significantly disagreeing  with observations is another indication that  the {\it hidden} scale symmetry is manifested in the density regime of compact stars.
\subsection{The PREX-II ``Dilemma"\\  and Hadron-Quark Duality}
We are now equipped with what enables us to address the PREX-II dilemma and the issue of  whether in EFT the EoS at low density near $n_0$ must {\it necessarily} constrain what happens at higher densities relevant to massive stars.

One can read off from the HDsF-driven  $E_{sym}$ ( Fig.~\ref{EsymGnEFT}) that $J\approx 30.2$ MeV and $L\approx 67.8$ MeV.  $J$ is ``soft" consistent with (\ref{accepted}) but the slope $L$ comes higher than the central value of (\ref{accepted}) by $\sim 10$ MeV.  Reliable S$\chi$EFT calculations to N$^3$LO converge to the central value of $\sim (52-56)$ MeV~\cite{schieft} which is consistent with (\ref{accepted}). 

What does this difference of $\sim 10$ MeV mean? 

This can be understood as that the ``soft" EoS at $n\lsim n_0$ starts to stiffen as the density approaches $n_{1/2}\gsim 2n_0$ reflecting the cusp smoothed by the HDsF.  This leads to $\Esym (2n_0)\approx 56.4$ MeV consistent with what is indicated in nature~\cite{BAL}. This reflects  that the S$\chi$EFT defined with the cutoff $\Lambda \lsim m_V$ starts breaking down at $\sim 2n_0$ precisely due to the {\it necessity} of the HDsF signaling the emergence -- vial Seiberg-type duality -- of QCD degrees of freedom. This crossover not only accounts for the massive star masses but also provides a simple mechanism to bring -- with additional help with the crust consistently treated thermodynamically -- $\Lambda_{1.4}\sim 650$ (which is still consistent with the presently accepted with the upper bound)  to a lower value in the vicinity of $\sim 400$ which may be favored should the tidal deformability bound be further tightened in the future measurements. Again there is a logically simple reason for this. The heavy-meson-induced smoothing tends to locate the central density of  the $\sim 1.4 M_\odot$ star for $\Lambda_{1.4}$  in the density regime  $< n_{1/2}$, i.e., the  lower side of  the cusp -- which is soft -- that could be in principle accurately calculated by S$\chi$EFTs (and $Gn$EFT) by fine-tuning the  crossover density $n_{1/2}$ within the range involved. 

We are then led to suggest that the strong $R_{\rm skin}$-$L$ correlation in the PREX-II measurement could not directly constrain the EoS of the core of compact stars. There is, in fact, nothing special about arriving at this sort of conclusion in {\it effective field theories} for QCD.   An apt example, perhaps not widely recognized in nuclear physics community,  is the applicability of the skyrmion approach -- as an EFT -- to nuclear physics. Given that the skyrmion theory should be a good low-energy effective field theory of QCD at large $N_c$ limit, it should in principle describe various different low-energy properties of nuclear physics valid at large $N_c$. Indeed in some cases, it works extremely well. For instance,  the BPS skyrmion is seen to give an excellent description of nuclear binding energies~\cite{bps} and radii~\cite{radii} for a wide range of nuclei  from light to heavy nuclei  $A >  200$. But the same BPS Lagrangian by itself does not satisfy the soft-pion theorems, the hall-mark of the current algebra and chiral symmetry.  This seems at odds with the general belief that nuclear phenomena are governed by chiral symmetry which in the skyrmion theory is encoded in the current algebra term in the Lagrangian. But it does not necessarily mean that soft theorems, naively interpreted, must give the constraint in the domain where the BPS structure is more appropriate. It is now understood that the infinite tower of vector mesons, say,  HDsF generalized from what we have been discussing, subsumed in the BPS Lagrangian is at work for the particular nuclear properties concerned, including cluster phenomenon in light nuclei~\cite{sutcliffe}.  

Furthermore one can write down~\cite{bps-sum}  a skyrmion model as a sum of BPS submodels, each of which has its own characteristics applicable to different regions of scales and dynamics. How to go from one to others is of course an open issue that remains to be clarified. It is clear, however, that it is not necessarily constrained by nLORE.  The recent discovery of Seiberg-type dualities for sHLS~\cite{komargodski, karasik,Y,kitano-matsudo} indicates that there may intervene more than the skyrmion-half-skyrmion topology change we have been exploiting in the phase structure of dense hadronic matter in going to high densities in the core of massive stars, a notable current example being the phase where the $\eta^\prime$ ring singularity is exposed, say,  in the vicinity of the putative IR fixed point~ density~\cite{dichotomy}.  Such a multiple phase structure involving ``hadrons" in place of quarks and gluons could persist all the way to the density where the hadron-quark continuity does truly break down~\cite{breakdown-HQ-continuity}.
\subsection{Strangeness Plays No Role}

 In what's treated in this paper and elsewhere, the strangeness degrees of freedom played no role. This seems at odds with the dilaton scheme we are adopting~\cite{GD} where kaons figure on the same footing as  the dilaton.  This point is discussed in \cite{MR-review}. It turns out however to be feasible to argue that the so-called ``hyperon problem" is absent in the density regime relevant to the core of massive neutron stars. The argument was based on the RG approach to interacting protons and neutrons coupled to the HDsF and the kaons on the Fermi surface~\cite{WGP-MR}. Invoking the same large $N_c\to \infty$ and large $\bar{N}=\frac{k_F}{\Lambda - k_F}\to \infty$ limit that underlies the results obtained in this paper (and more generally~\cite{MR-review}), it was shown~\cite{MR-nostrangeness} that (1) kaons condense and hyperons appear at  about the same density and (2) the kaon condensation threshold density $n_K$ satisfies the bound 
\be
n_K > \bar{N} n_0.\label{Bound}
\ee
This implies that  $n_K$ could be considerably higher than  the core density of the stars.  We note that the bound (\ref{Bound}) with $n_K >7 n_0$ was arrived at in a different but related consideration -- short-range correlations --  a long time ago~\cite{panda}.  A rigorous justification for this could be given in the $V_{lowk}$ RG approach to $Gn$EFT $\in SU(3)_f$,  which remains to be worked out.

\section{Concluding remarks} 
Starting with the cusp structure found in a skyrmion-crystal simulation, with the incorporation of heavy degrees of freedom considered to be dual to the gluons/quarks in the EoS for dense matter, we have arrived at the symmetry energy $\Esym (n)$ that contraries the nLORE (standard nuclear astrophysics lore) and gives rise to the pseudo-conformal sound speed for $n > n_{1/2}$.  

Two strikingly novel features, while admittedly unorthodox and drastically different from all other predictions made in the field, that are -- up-to-date -- not in serious tension with all available experimental data are as follows.

$\bullet$ {\bf First: {\it The pseudo-conformal structure of the EoS found in this paper  is tied indispensably to the symmetry energy $E_{sym} (n)$ for $n\lsim 7n_0$  with the cusp at $n_{1/2} \sim (2-3) n_0$ encoding the topology change representing the hadron-quark duality.}}

$\bullet$ {\bf Second: {\it It follows straightforwardly -- and directly -- from the structure of $E_{sym}$ that the sound speed of massive stars becomes precociously pseudo-conformal in the $V_{lowk} RG$ approximation,  $v_{pcs}^2/c^2\approx 1/3$, for $n\gsim n_{1/2}$ and that the structure of the core populated by fractionally-charged quasi-fermions mocks up  surprisingly closely that of deconfined quarks~\cite{evidence}. }}

Any confirmed evidence that these features are falsified by Nature (or by rigorous theories) would torpedo  this approach. There are many papers in the literature, too many to cite\footnote{Just to  give a few recent sources for tracing back, we cite  \cite{sources}.},  that discuss how one can ``derive" or deduce the speed of sound from observables and vice-versa.  Our approach is seemingly at odds with this claim.  This was pointed out in Sect.~\ref{topology-encoded} in some of the cases we have looked at. What leads crucially to the pseudo-conformal structure of the sound speed in our theory seems to be predominantly controlled by the emergent (albeit approximate) scale symmetry combined with the vector manifestation of the hidden gauge coupling at $n\gsim 25 n_0$~\cite{PKLMR}. The perhaps deceptively ``simple" structure of the high-density property of the trace of the energy-momentum tensor,  going proportional to $\la\chi\ra^4$ within the star density range with the dilaton condensate proportional to the chiral invariant mass $m_0$~\cite{PKLMR},   seems  to leave unaffected appreciably  the EoS that controls the star properties so far studied. 

This ``unorthodox" feature of the EoS may have something to do with the highly controversial, poorly understood, nature of scale or conformal symmetry in strong interactions.  
There is an issue as to whether the dilaton that figures in the pseudo-conformal structure in our approach is connected to the would-be QCD dilaton GD~\cite{GD} or to the CD with the (conjectured) IR fixed point in the conformal phase~\cite{DDZ}.  Our speculation is that the hidden scale symmetry emerging in nuclear and compact-star physics brings the CD (conformal dilaton) to coincide with the GD (genuine dilaton) .

\subsection*{Acknowledgments}
The work of YLM was supported in part by the National Science Foundation of China (NSFC) under Grant No. 11875147 and 11475071.

\end{document}